\documentclass{article}

\usepackage{arxiv}

\usepackage[utf8]{inputenc} 
\usepackage[T1]{fontenc}    
\usepackage{hyperref}       
\usepackage{url}            
\usepackage{booktabs}       
\usepackage{amsfonts}       
\usepackage{nicefrac}       
\usepackage{microtype}      
\usepackage{lipsum}		
\usepackage{graphicx}
\usepackage{natbib}
\usepackage{doi}
\usepackage{amsmath}

\usepackage{fancyhdr}
\usepackage{textcomp}

\title{Tacit Signal Infrastructure: Towards AI Systems that Model Expert Sensing Over Time}

\author{ 
	\href{https://orcid.org/0009-0004-1760-0149}{\includegraphics[scale=0.06]{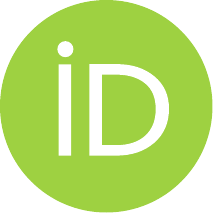}\hspace{1mm}Annie Yuan} \\
	School of Computer Science\\
	The University of Sydney\\
	NSW, 2006, Australia \\
	\texttt{annie.yuan@sydney.edu.au} \\
}

\renewcommand{\shorttitle}{Tacit Signal Infrastructure}

\hypersetup{
pdftitle={Tacit Signal Infrastructure: Towards AI Systems that Model Expert Sensing Over Time},
pdfsubject={Artificial Intelligence, Human--AI Interaction, Expert Cognition, Cognitive Infrastructure},
pdfauthor={Annie Yuan},
pdfkeywords={Tacit Signal Infrastructure, Expert Weak Signal Perception, Weak Cognitive Signals, Tacit Signal Computability, Longitudinal Cognitive Operations, Expert Cognition, Weak Signals}
}

\begin{document}
\maketitle
\thispagestyle{plain}

\begin{abstract}
Current generative AI systems are increasingly effective at processing explicit knowledge: retrieving information, summarising documents, generating explanations, and supporting codified workflows. However, high-level expertise is not reducible to explicit knowledge. Human experts rely on tacit sensing: perceiving weak signals, recognising emerging tensions, detecting coherence degradation, and anticipating instability before formal indicators appear. Existing AI education, AI literacy, and human--AI collaboration frameworks remain centred on prompting, task execution, and productivity support, and are poorly equipped to address this tacit layer of expert cognition.
This vision paper argues that next-generation AI systems should move beyond explicit knowledge processing toward the longitudinal modelling of expert tacit sensing. It introduces \textit{Tacit Signal Infrastructure} as a layer for supporting the capture, structuring, modelling, and interpretation of expert tacit signals over time. It further defines \textit{Long-term Cognitive Operations} as the practices required to maintain, validate, and govern AI systems that model expert cognition beyond explicit knowledge, including memory curation, semantic organisation, tacit signal modelling, reasoning calibration, and cognitive governance.
Building on this framing, the paper proposes the \textit{Cognitive Operations Manager} as a prototype AI-native professional role for coordinating tacit signal capture and modelling, semantic modelling, AI system calibration, expert validation, and ethical governance. It also introduces a \textit{Cognitive Operations Research and Training Framework} (CORTF) to support research, education, and workforce development. The paper contributes a conceptual foundation for designing AI systems that model expert sensing over time, while positioning cognition as an infrastructural, operational, and professional domain in persistent human--AI systems.
\end{abstract}


\section{Introduction}

Generative AI is shifting the role of artificial intelligence in professional work. Earlier digital systems primarily supported computation, information retrieval, communication, and workflow automation. More recent generative AI systems increasingly participate in activities closer to cognition itself: reasoning with users, organising knowledge, generating explanations, maintaining conversational context, supporting decision-making, and assisting with expertise development. This shift suggests that AI should no longer be understood only as a tool for isolated task execution, but also as part of a persistent cognitive environment in which human and machine capabilities become increasingly interdependent.

At the same time, current generative AI systems remain strongest when working with explicit knowledge. They can retrieve information, summarise documents, generate procedural explanations, transform codified workflows, and support semantic organisation across large volumes of text. These capabilities are powerful, but they also reveal a deeper limitation. High-level expertise is not reducible to explicit knowledge. Human experts often rely on tacit forms of judgement: perceiving weak signals, recognising emerging tensions, sensing coherence degradation, and anticipating instability before formal indicators appear. Theories of tacit knowledge, expert intuition, naturalistic decision-making, and reflective professional practice have long shown that expert cognition often involves forms of knowing that are difficult to fully articulate or formalise \cite{polanyi1966logic,dreyfus1986mind,klein2017sources,schon2017reflective}.

This paper argues that this tacit layer of expert cognition is becoming increasingly important in the AI era. As explicit knowledge is progressively externalised into machine-readable systems, the distinctive value of expert professionals may increasingly lie not only in possessing information, but in sensing what is not yet fully visible: subtle shifts in meaning, rhythm, coherence, tension, trust, direction, or systemic stability. These weak signals are often not captured in formal documentation, task logs, performance indicators, or short-term interaction traces. Yet they frequently shape expert judgement in complex professional environments.

The difficulty of eliciting and formalising expert knowledge has long been recognised in knowledge engineering, particularly where expert judgement depends on perceptual distinctions, contextual cues, and tacit reasoning processes that experts may not be able to fully verbalise \cite{hoffman1987problem}. This challenge becomes more significant for next-generation AI systems. If AI systems are to move beyond explicit knowledge processing and model expert cognition more deeply, they must address not only what experts know, but how experts sense, interpret, and anticipate change across time.

Much existing research and practice still frames AI use at the level of interaction. Human--AI interaction research has produced important principles for designing systems that are understandable, controllable, and responsive to human correction \cite{amershi2019guidelines,shneiderman2022human}. Similarly, current approaches to AI literacy and AI education often emphasise responsible use, evaluation of AI outputs, effective prompting, task completion, and productivity support \cite{kasneci2023chatgpt}. These perspectives remain valuable, but they are insufficient for understanding how AI systems might model expert cognition beyond explicit knowledge, especially when tacit sensing unfolds longitudinally across time.

The central issue is therefore a change in the unit of analysis. If AI systems are to model expert sensing, they cannot rely only on isolated prompts, explicit knowledge bases, or short-term task outcomes. Because tacit signals are rarely visible in isolated interactions, such systems require persistent infrastructures capable of capturing cognitive traces across time, organising semantic structures, modelling weak signals, and supporting expert feedback on whether emerging interpretations are meaningful. In such settings, cognition is not confined to an individual mind or a discrete interface interaction. It becomes distributed across people, AI systems, memory stores, documents, semantic networks, workflows, and professional routines.

This view is consistent with long-standing theories of distributed and extended cognition, which argue that cognitive activity can be organised across people, artefacts, tools, and environments rather than located solely within the individual \cite{hutchins1995cognition,clark1998extended,hollan2000distributed}. It also resonates with infrastructure studies, which show that infrastructures are often most powerful when they become embedded, persistent, and taken for granted within everyday practice \cite{star1999ethnography}. From this perspective, AI systems that aim to model expert cognition beyond explicit knowledge require more than improved interfaces or larger knowledge bases; they require infrastructures capable of capturing, structuring, and validating tacit signals as they unfold over time.

Existing frameworks do not yet fully address this transition. Human--AI interaction research has largely focused on designing better interactions between users and intelligent systems. AI literacy has largely focused on helping individuals understand, evaluate, and use AI responsibly. Knowledge management and organisational learning have examined how organisations create, store, and share knowledge \cite{argyris1996organizational,nonaka2007knowledge}, but they were not developed for environments in which AI systems attempt to model tacit expert sensing over time. As a result, there remains a conceptual gap in understanding how expert cognition beyond explicit knowledge can be captured, interpreted, validated, and governed within persistent AI systems.

To address this gap, this paper introduces the concept of \textit{Tacit Signal Infrastructure}. Tacit Signal Infrastructure refers to a cognitive infrastructure layer for supporting the capture, structuring, modelling, interpretation, and validation of expert tacit signals over time. Such signals may include patterns of coherence degradation, emerging tension, rhythm disruption, semantic instability, value conflict, or anticipatory risk. The aim is not to fully automate expert judgement, but to develop infrastructures that make tacit signals more visible, reviewable, and usable within next-generation AI systems.

The theoretical chain developed in this paper begins with \textit{human tacit cognition}: forms of expert judgement that are difficult to fully articulate or formalise. Within this tacit layer, experts perceive \textit{weak cognitive signals}, including emerging tensions, coherence degradation, rhythm disruption, semantic instability, and anticipatory instability. This capability is defined here as \textit{expert weak-signal perception}. The paper then proposes \textit{Tacit Signal Infrastructure} as a means of capturing, structuring, modelling, and validating such signals over time. Through \textit{longitudinal weak-signal calibration}, tacit signals may become partially computable without being reduced to fixed rules. This provides a foundation for AI-supported expert cognition and motivates \textit{Long-term Cognitive Operations} as the broader operational paradigm required to maintain and govern such systems.

This paper further introduces \textit{Long-term Cognitive Operations} as the operational paradigm required to maintain, validate, and govern AI systems that model expert cognition beyond explicit knowledge. Long-term Cognitive Operations includes memory curation, semantic organisation, tacit signal modelling, reasoning calibration, expertise development, role calibration, and cognitive governance. Together, Tacit Signal Infrastructure and Long-term Cognitive Operations provide a framework for designing AI systems that support the longitudinal modelling of expert sensing while preserving the need for human interpretation, contestability, and ethical oversight.

The paper also argues that these systems may give rise to new forms of AI-native professional capability. As cognitive infrastructures become more complex, organisations may require professionals who can coordinate tacit signal capture and modelling, semantic modelling, AI system calibration, expert feedback, and ethical governance. This paper introduces the \textit{Cognitive Operations Manager} (COM) as a prototype professional role for understanding the capabilities that may become necessary when AI systems move beyond explicit knowledge processing toward the modelling of expert cognition.

Finally, the paper proposes a \textit{Cognitive Operations Research and Training Framework} (CORTF) as a preliminary architecture for research, education, and workforce development. The purpose of CORTF is to support the systematic study and teaching of long-term cognitive operations, including how professionals can design, evaluate, govern, and operate next-generation AI cognitive systems.

This vision paper makes five contributions. First, it identifies a limitation in current AI systems: while they increasingly process explicit knowledge effectively, they remain poorly equipped to model the tacit sensing abilities through which human experts perceive weak cognitive signals, emerging tensions, coherence degradation, and anticipatory instability. Second, it introduces \textit{Tacit Signal Infrastructure} as a cognitive infrastructure layer for supporting the capture, structuring, modelling, interpretation, and validation of expert tacit signals over time. Third, it defines \textit{Long-term Cognitive Operations} as the operational paradigm required to maintain, validate, and govern AI systems that model expert cognition beyond explicit knowledge. Fourth, it proposes the \textit{Cognitive Operations Manager} as a prototype professional role responsible for coordinating tacit signal modelling, semantic modelling, AI system calibration, expert feedback, and ethical governance. Fifth, it presents the \textit{Cognitive Operations Research and Training Framework} as a research and educational architecture for developing the capabilities required to design, evaluate, and operate next-generation AI cognitive systems.

\section{Related Work}

This paper builds on several research traditions concerned with expert cognition, weak-signal interpretation, human--AI interaction, distributed cognition, infrastructure, and organisational knowledge. These areas provide important foundations for understanding why expert judgement cannot be reduced to explicit knowledge and why future AI systems may require persistent cognitive infrastructures. However, existing work has not yet fully addressed the problem central to this paper: how AI systems might model expert tacit sensing longitudinally through dedicated Tacit Signal Infrastructure.

\subsection{Tacit Knowledge, Expertise, and Expert Judgement}

Research on tacit knowledge has long argued that expert knowing cannot be fully reduced to explicit rules or formal representations. Polanyi's account of tacit inference emphasises that people often know more than they can explicitly state, while work on expert performance has shown that expertise frequently involves situated judgement, embodied skill, and context-sensitive action rather than rule application alone \cite{polanyi1966logic,dreyfus1986mind}. Naturalistic decision-making further demonstrates that experts often act by recognising meaningful patterns in complex environments, especially under uncertainty and time pressure \cite{klein2017sources}. Similarly, reflective practice highlights how professionals think in action, adapting their judgement within uncertain and evolving situations \cite{schon2017reflective}.

These traditions are directly relevant to the present paper because tacit sensing is treated as a core dimension of expert cognition. The difficulty of extracting expert knowledge has also been recognised in knowledge engineering, particularly when expertise depends on perceptual distinctions, contextual cues, and forms of judgement that experts may struggle to verbalise \cite{hoffman1987problem}. However, existing work has not yet fully examined how AI systems might support the longitudinal modelling of such tacit signals. This paper extends these foundations by proposing Tacit Signal Infrastructure as a way to make expert weak-signal perception more visible, reviewable, and computationally usable without reducing it to fixed rules.

\subsection{Weak Signals, Early Warning Signs, and Anticipatory Expertise}

The concept of weak signals has long been used in strategic management and foresight to describe early, incomplete, and ambiguous indications of future change or strategic surprise \cite{ansoff1975managing}. Subsequent work has further developed weak signals as future-oriented signs that require interpretation, contextualisation, and attention before they become strong or institutionally recognised \cite{hiltunen2008future}. Research on weak-signal filters also suggests that organisations may fail to notice or act on such signals when they conflict with existing assumptions, routines, or decision structures \cite{ilmola2006filters}.

A related project management literature examines early warning signs in complex projects, where emerging problems may become visible before formal project failure occurs \cite{williams2012identifying,hajikazemi2013review}. This work supports the view that professionals must attend not only to formal metrics, but also to subtle indications of future difficulty. Research on situation awareness similarly conceptualises expert performance as involving the perception of relevant cues, comprehension of their meaning, and projection of future states \cite{endsley1995toward}. Work on high-reliability organisations further emphasises sensitivity to operations and attention to small signs of potential failure before breakdown occurs \cite{weick2011managing}.

These literatures are important because they show that weak-signal perception is already recognised as valuable in strategic, project, and high-stakes operational contexts. However, weak-signal research has generally focused on organisational foresight, risk detection, or early warning processes rather than on modelling weak-signal perception as a form of tacit expert cognition. This paper connects weak signals to tacit expertise and argues that future AI systems require infrastructures capable of modelling how experts sense, interpret, and validate such signals over time.

\subsection{Human--AI Interaction, Sensemaking, and Tacit Knowledge Elicitation}

Human--AI interaction research provides important principles for designing AI systems that are intelligible, controllable, correctable, and aligned with human needs. Guidelines for human--AI interaction emphasise transparency, appropriate reliance, user control, and mechanisms for correction, while human-centred AI argues for systems that amplify human capability while preserving human agency and responsibility \cite{amershi2019guidelines,shneiderman2022human}. Explanation research similarly highlights that AI explanations should support human understanding, contestability, and decision-making rather than merely expose technical model details \cite{miller2019explanation}.

HCI and CSCW research has also begun to examine tacit knowledge in professional and computational work. Studies of machine learning and data work show that practitioners rely on contextual judgement, informal know-how, and situated expertise that are not easily captured by formal workflows or documentation \cite{cha2023unlocking}. Related work on intelligent assistants for tacit knowledge elicitation suggests that computational systems may support the articulation and transfer of otherwise informal professional knowledge \cite{freire2023tacit}. Sensemaking and visual analytics research has also explored how computational tools can support analysts in organising information, detecting patterns, and constructing interpretations in complex information environments \cite{pirolli2005sensemaking}.

These studies show that HCI has recognised tacit knowledge, sensemaking, and human judgement as important concerns in computational systems. However, existing work has primarily focused on supporting human interpretation, eliciting tacit knowledge, or improving interaction with AI tools. Less attention has been given to tacit sensing as a longitudinal expert capability: the ability to perceive weak signals, emerging tensions, coherence degradation, and anticipatory instability before these phenomena become formally measurable. This paper therefore extends HCI work by framing weak-signal perception as an infrastructural problem for future AI systems.

\subsection{Distributed Cognition, Infrastructure, and Organisational Knowledge}

The proposed framework also draws on theories of distributed and extended cognition. These theories challenge the view that cognition is located solely within the individual mind, instead showing how cognitive activity can be organised across people, artefacts, tools, representations, and environments \cite{hutchins1995cognition,clark1998extended,hollan2000distributed}. This perspective is important for understanding AI systems not merely as external tools, but as components within broader cognitive arrangements.

Infrastructure studies further show that infrastructures shape practice by becoming embedded, persistent, and often taken for granted within everyday activity \cite{star1999ethnography}. Classification and infrastructure work also demonstrates that representational systems do not merely store information; they structure what can be seen, organised, and acted upon \cite{bowker2000sorting}. From this perspective, future AI systems may function not only as interfaces for task execution, but as cognitive infrastructures that organise memory, reasoning, semantic structures, and expert interpretation over time.

Knowledge management and organisational learning research has examined how organisations create, store, share, and transform knowledge. Organisational learning theory highlights how organisations detect and correct errors, reflect on practice, and adapt their routines over time \cite{argyris1996organizational}. The knowledge creation literature has also emphasised the relationship between tacit and explicit knowledge, particularly how tacit knowledge may be externalised, shared, and recombined within organisational contexts \cite{nonaka2007knowledge}.

These traditions are highly relevant to the proposed framework, but they were not developed for persistent AI systems that attempt to model tacit expert sensing over time. Existing knowledge management approaches primarily focus on organisational knowledge as something to be created, stored, shared, and applied. By contrast, Long-term Cognitive Operations treats cognition itself as an operational domain requiring ongoing memory curation, semantic orchestration, tacit signal validation, reasoning calibration, and governance. This shift motivates the need for new professional capabilities, such as the Cognitive Operations Manager, to coordinate cognitive infrastructures in AI-mediated professional environments.

\section{From Explicit Knowledge Systems to Tacit Signal Infrastructure}

Having established the limits of interaction-centred AI and explicit knowledge processing, this section develops the central construct of the paper: Tacit Signal Infrastructure. It argues that modelling expert cognition requires attention to weakly articulated, temporally situated signals that are not adequately represented in conventional knowledge systems.

However, expert cognition is not exhausted by explicit knowledge. In many professional domains, experienced practitioners do not only apply rules, procedures, or documented knowledge. They also perceive subtle signals within complex situations, recognise emerging instability, detect incoherence before it becomes measurable, and make anticipatory judgements under uncertainty. These forms of judgement are often difficult to formalise because they are grounded in experience, context, pattern recognition, and tacit sensitivity to changing conditions.

This section establishes the conceptual transition from explicit knowledge systems to tacit signal infrastructure. It argues that if future AI systems are to model expert cognition more deeply, they must move beyond explicit semantic knowledge and develop mechanisms for capturing, structuring, and interpreting tacit signals over time.

\subsection{The Limits of Explicit Knowledge Systems}

Most contemporary professional information systems are built around explicit knowledge. They manage documents, procedures, rules, workflows, performance metrics, databases, taxonomies, and institutional knowledge repositories. These infrastructures are essential for professional coordination because they allow knowledge to be stored, transmitted, searched, standardised, and audited.

Generative AI extends these explicit knowledge systems by making them more interactive and generative. Instead of merely retrieving documents, AI systems can summarise, synthesise, translate, classify, and reconfigure information in response to user goals. This enables new forms of productivity and knowledge access. Yet the underlying orientation remains largely explicit: the system works with information that has already been expressed, recorded, or made available as data.

The limitation is that many important forms of expert judgement are not initially available in explicit form. Traditional professional infrastructures often assume that expertise can be sufficiently represented through rules, procedures, technical documentation, standardised workflows, measurable performance indicators, and formal decision frameworks. However, experienced professionals frequently detect meaningful changes before they are visible in these explicit systems. They may sense organisational instability before measurable decline, detect project incoherence before operational failure, perceive emotional fragmentation within teams before visible conflict, or identify long-term systemic risk despite short-term numerical stability.

These forms of perception are rarely reducible to formal metrics. Experts often describe them through intuitive or metaphorical language, such as sensing that a project is losing coherence, that a team has lost direction, that communication has become mechanical, or that a system no longer feels aligned. Such descriptions indicate that expert cognition frequently operates through weakly articulated perceptual structures rather than explicit analytical reasoning alone.

In this sense, explicit knowledge systems are necessary but insufficient for modelling expert cognition. They can represent what is already codified, but they struggle to capture how experts sense, interpret, and anticipate change before it becomes formally visible.

\subsection{Tacit Sensing as an Expert Capability}

This paper uses the term \textit{tacit sensing} to refer to the expert ability to perceive weak, ambiguous, or weakly articulated signals within evolving systems. Tacit sensing is not merely intuition in a vague sense. Rather, it refers to situated perceptual and interpretive capabilities developed through sustained engagement with a domain.

\begin{quote}
\textbf{Definition.} Tacit sensing refers to the expert capability to perceive weak signals, emerging tensions, coherence degradation, rhythm disruption, value conflict, or anticipatory instability before these phenomena become fully explicit or formally measurable.
\end{quote}

Tacit sensing often emerges through prolonged experiential exposure, repeated interaction with complex systems, temporal pattern recognition, embodied judgement, and long-term adaptive observation. It is one of the ways in which experts operate not only as information processors, but as temporal sensing systems capable of anticipating instability before formal breakdown occurs.

In professional environments, weak signals may include shifts in communicative tone, declining creative spontaneity, behavioural rigidity, emotional disengagement, fragmented collaboration, slowing decision rhythms, or subtle incoherence between strategic narratives and operational behaviour. These signals are often recognised before explicit evidence becomes available. Their significance depends not only on their presence, but on how they are interpreted in relation to prior experience, domain knowledge, and unfolding trajectories.

Tacit sensing is professionally valuable because it enables early intervention. Experts can adjust direction, reframe a problem, support a learner, redesign a process, or question an assumption before failure becomes visible. Tacit sensing therefore has a temporal structure: it concerns not only what is happening now, but what appears to be emerging.

\subsection{Micro-Cognitive Signals and Professional Differentiation}

Although tacit sensing appears across professions, its expression is domain-specific. Different expert communities develop distinct vocabularies for describing instability, coherence, rhythm, and systemic imbalance. For example, a design director may describe a project as losing coherence; a film director may sense that performative flow has disappeared; a meditation instructor may perceive attention fragmentation; and an organisational leader may identify mechanical communication patterns within a team.

Despite these linguistic differences, such descriptions may point toward structurally similar cognitive phenomena, including coherence degradation, rhythm disruption, anticipatory instability detection, semantic fragmentation, and systemic misalignment. This suggests that professional expertise contains a hidden layer of micro-cognitive signals operating beneath formal occupational language.

This paper uses the term \textit{micro-cognitive structures} to describe the fine-grained perceptual and interpretive patterns through which experts perceive and organise such signals.

\begin{quote}
\textbf{Definition.} Micro-cognitive structures are fine-grained perceptual and interpretive patterns through which experts recognise weak signals, emerging tensions, coherence shifts, rhythm disruptions, and anticipatory changes within complex professional situations.
\end{quote}

Micro-cognitive structures are important because they provide a bridge between raw behaviour and expert judgement. For example, repeated hesitation may be interpreted as uncertainty, but in a particular context it may also signal conceptual conflict, lack of trust, role ambiguity, or misalignment between values and action. The meaning of the signal depends on expert interpretation.

For AI systems, this creates a major challenge. Weak signals cannot be treated as simple variables detached from context. They must be interpreted longitudinally, semantically, and relationally. This is why tacit sensing cannot be modelled adequately through isolated interaction logs or explicit knowledge bases alone.

\subsection{Temporal Anticipation and Longitudinal Cognition}

A further limitation of conventional professional systems lies in their static treatment of cognition. Most educational and organisational infrastructures evaluate individuals through episodic snapshots, such as examinations, periodic reviews, isolated competency assessments, project milestones, or performance metrics. However, expert cognition is fundamentally temporal.

High-level professionals do not merely respond to present conditions. They continuously interpret trajectories, anticipate future instability, and adjust behaviour through longitudinal perception. This temporal dimension is especially important in complex environments where organisational structures evolve continuously, project ecosystems remain unstable, interdisciplinary coordination changes dynamically, and AI systems increasingly participate in decision environments.

Under such conditions, cognition becomes an ongoing operational process rather than a fixed knowledge state. Future AI systems that aim to model expert sensing therefore require mechanisms for continuously capturing cognition over time, tracking evolving decision structures, identifying recurring tacit patterns, modelling long-term value orientations, and detecting emerging weak signals across professional trajectories.

This establishes the necessity for Long-term Cognitive Operations: the operational paradigm through which cognitive infrastructures can be maintained, validated, and governed over time.

\subsection{Tacit Signal Infrastructure}

To support next-generation AI systems that model expert cognition beyond explicit knowledge, this paper introduces the concept of \textit{Tacit Signal Infrastructure}.

\begin{quote}
\textbf{Definition.} Tacit Signal Infrastructure refers to a cognitive infrastructure layer that supports the capture, structuring, interpretation, and validation of expert tacit signals over time, enabling AI systems to model weak-signal perception, anticipatory judgement, and expert sensing beyond explicit knowledge.
\end{quote}

Tacit Signal Infrastructure is not a single dashboard, dataset, or algorithm. It is a layered infrastructure that connects multiple processes: capturing professional activity, organising semantic knowledge, extracting potential weak signals, modelling cognition longitudinally, supporting AI Twin reasoning, and enabling analytics across time.

The purpose of this infrastructure is not to fully automate expert judgement. Such a goal would risk oversimplifying or distorting tacit cognition. Instead, Tacit Signal Infrastructure aims to make tacit signals more visible, reviewable, and computationally usable. It supports AI systems in learning from expert sensing while preserving the need for human interpretation, contestability, and contextual judgement.

Tacit Signal Infrastructure therefore extends explicit knowledge systems in three ways. First, it shifts attention from codified knowledge to weakly articulated signals. Second, it introduces longitudinal modelling, because tacit signals often become meaningful only across time. Third, it requires human validation, because expert judgement cannot be reduced to signal extraction alone.

\section{Longitudinal Cognition and Long-term Cognitive Operations}

Tacit signals are rarely meaningful as isolated events. A single hesitation, inconsistency, or shift in tone may have many possible explanations. What makes such signals significant is often their recurrence, trajectory, context, and relation to other signals over time. For this reason, modelling expert tacit sensing requires a longitudinal view of cognition.

This section introduces \textit{Longitudinal Cognition} and \textit{Long-term Cognitive Operations}. Longitudinal Cognition describes the temporal evolution of cognitive states, semantic structures, tacit signals, and expertise trajectories. Long-term Cognitive Operations describes the practices required to maintain, validate, and govern these evolving cognitive infrastructures.

\subsection{Longitudinal Cognition}

\begin{quote}
\textbf{Definition.} Longitudinal Cognition refers to the continuous evolution of human--AI cognitive states, memory structures, semantic systems, tacit signals, and expertise trajectories across time.
\end{quote}

Longitudinal Cognition differs from interaction-based models of AI use. Interaction-based models focus on what occurs within a discrete exchange: a prompt, a response, a task, or a session. Longitudinal Cognition instead asks how cognition develops across repeated interactions, accumulated memories, evolving semantic structures, and changing professional contexts.

This is especially important for tacit sensing. Expert perception is not formed through one-off observation. It develops through repeated exposure to patterns, anomalies, tensions, breakdowns, and recoveries. Experts learn what matters by experiencing how systems change over time. Similarly, AI systems that aim to model expert sensing must be able to connect present signals with prior trajectories.

Longitudinal Cognition therefore involves several dimensions. The first is \textit{memory continuity}, where prior interactions, decisions, and observations remain available for future interpretation. The second is \textit{semantic accumulation}, where concepts, categories, and relationships become more refined over time. The third is \textit{tacit signal trajectory}, where weak signals are tracked across repeated contexts. The fourth is \textit{expertise evolution}, where human and AI systems develop more sophisticated interpretive capabilities through ongoing engagement.

\subsection{From Interaction to Operation}

The shift from interaction to operation is central to this paper. If AI systems are used only for isolated tasks, then the primary design challenge is interaction quality. However, if AI systems participate in the longitudinal modelling of expert cognition, the design challenge becomes operational: how to maintain memory, validate tacit signals, calibrate reasoning, organise semantic structures, and govern cognitive infrastructures over time.

\begin{table}[h]
\centering
\caption{From interaction-level AI use to Long-term Cognitive Operations.}
\begin{tabular}{p{0.28\linewidth}p{0.30\linewidth}p{0.30\linewidth}}
\hline
\textbf{Dimension} & \textbf{Interaction Paradigm} & \textbf{Operations Paradigm} \\
\hline
Temporal structure & Episodic sessions & Longitudinal trajectories \\
Primary object & Prompts and outputs & Cognitive infrastructure \\
Knowledge focus & Explicit information & Explicit knowledge and tacit signals \\
System behaviour & Reactive assistance & Persistent modelling and calibration \\
Human role & AI user & Cognitive system operator and validator \\
Evaluation focus & Output usefulness & Long-term cognitive integrity \\
\hline
\end{tabular}
\end{table}

In the operations paradigm, AI capability is not judged only by whether it produces useful outputs. It is also judged by whether the cognitive infrastructure remains coherent, trustworthy, adaptable, and aligned with expert judgement over time.

\subsection{Defining Long-term Cognitive Operations}

\begin{quote}
\textbf{Definition.} Long-term Cognitive Operations refers to the ongoing orchestration, validation, governance, and improvement of persistent human--AI cognitive infrastructures, including memory systems, semantic structures, tacit signal layers, reasoning processes, expertise models, and ethical safeguards.
\end{quote}

Long-term Cognitive Operations are not equivalent to prompt engineering. Prompting is concerned with how to elicit useful outputs from AI systems in particular moments. Long-term Cognitive Operations are concerned with how to sustain the quality of cognition across time.

They are also not equivalent to conventional workflow automation. Workflow automation focuses on task execution. Long-term Cognitive Operations focus on the maintenance of the cognitive conditions under which interpretation, judgement, and expertise development occur.

Finally, Long-term Cognitive Operations are not merely knowledge management. Knowledge management traditionally concerns the creation, storage, sharing, and use of organisational knowledge. Long-term Cognitive Operations extend this concern toward persistent AI systems that model memory, semantic structures, tacit signals, reasoning continuity, and expert judgement over time.

\subsection{Operational Domains}

Long-term Cognitive Operations can be organised into six interrelated operational domains.

\textit{Memory operations} concern the curation, retrieval, updating, and forgetting of persistent memory. This includes deciding what should be retained, how memory should be indexed, when prior information should influence current reasoning, and how outdated or harmful memory should be revised.

\textit{Semantic operations} concern the organisation and maintenance of meaning structures. This includes concept mapping, ontology development, terminology alignment, and the management of semantic consistency across documents, systems, agents, and professional contexts.

\textit{Tacit signal operations} concern the capture, interpretation, validation, and refinement of weak signals. This includes identifying possible indicators of coherence degradation, emerging tension, rhythm disruption, value conflict, or anticipatory instability, and validating these interpretations with human experts.

\textit{Reasoning operations} concern the calibration and supervision of AI-supported reasoning. This includes checking assumptions, identifying reasoning drift, evaluating inference quality, and ensuring that AI interpretations remain aligned with domain knowledge and expert judgement.

\textit{Expertise development operations} concern the modelling and support of human and AI expertise over time. This includes tracking how interpretive capabilities develop, identifying gaps in professional judgement, and designing systems that support reflective learning rather than replacing expertise.

\textit{Cognitive governance} concerns the ethical and organisational oversight of cognitive infrastructures. This includes privacy, consent, accountability, contestability, cognitive autonomy, and safeguards against surveillance, over-dependence, or inappropriate automation of judgement.

Together, these domains define Long-term Cognitive Operations as a new operational paradigm for AI systems that aim to model expert cognition beyond explicit knowledge.

\section{Layered Architecture for Longitudinal Cognitive Operations and Tacit Signal Infrastructure}

The modelling of tacit sensing, weak-signal perception, and longitudinal cognition transforms the role of AI infrastructure. Traditional professional systems primarily manage information, documentation, communication, and operational workflows. By contrast, next-generation cognitive infrastructures must support memory accumulation, tacit signal modelling, adaptive reasoning, long-term planning, and governance across time.

This section presents a layered architecture for Longitudinal Cognitive Operations and Tacit Signal Infrastructure. The architecture is not intended as a fixed technical specification. Rather, it provides a conceptual model for understanding how future AI systems may move from short-term response generation toward persistent cognitive infrastructures capable of modelling expert sensing over time.

\begin{figure}[t]
    \centering
    \includegraphics[width=\linewidth]{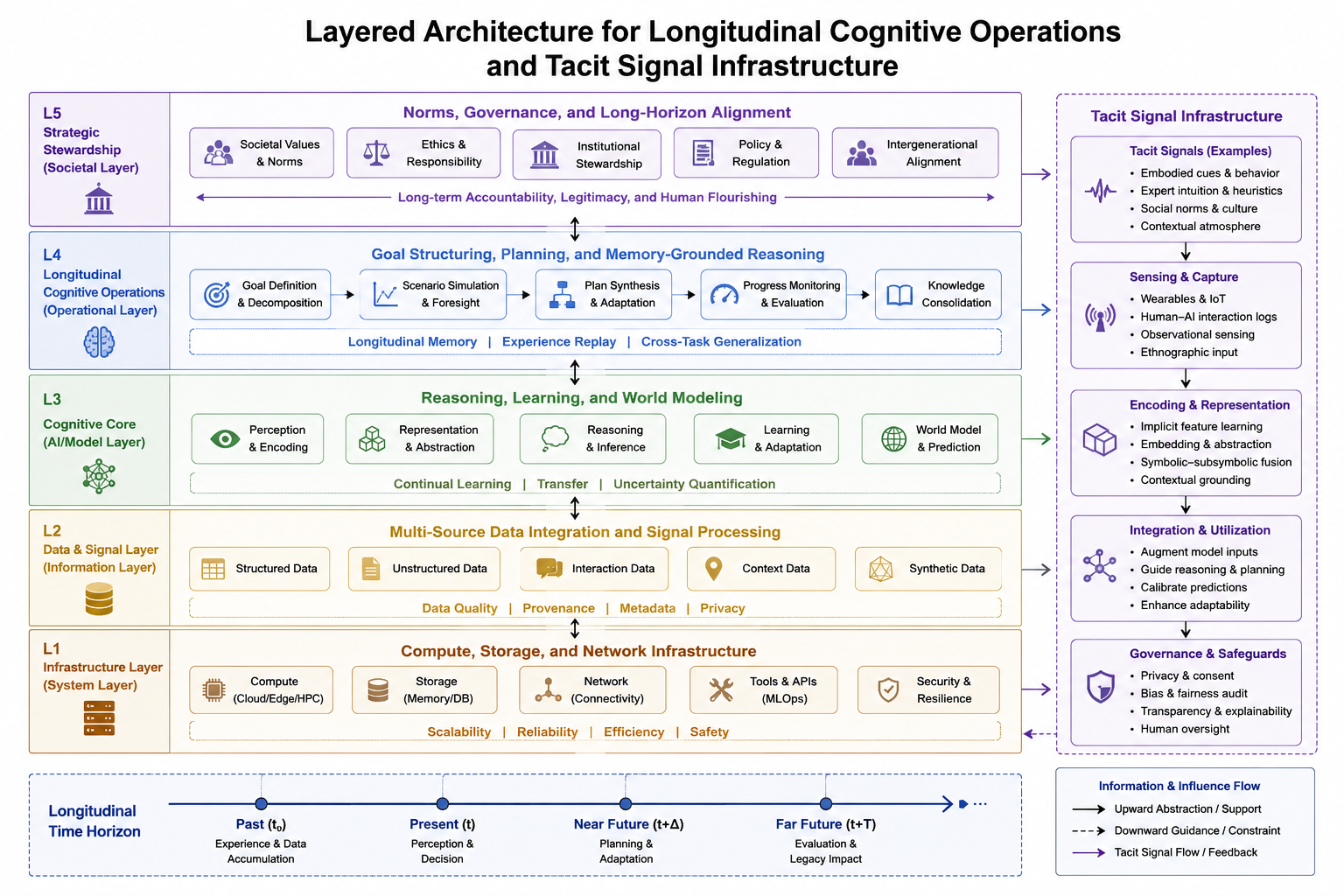}
    \caption{Layered architecture for Longitudinal Cognitive Operations and Tacit Signal Infrastructure.}
    \label{fig:longitudinal-cog-operations}
\end{figure}

Figure~\ref{fig:longitudinal-cog-operations} illustrates the proposed architecture. The framework conceptualises cognition not as a single interaction process, but as a longitudinal operational infrastructure integrating memory, reasoning, tacit signals, adaptive governance, and future-oriented planning. The architecture consists of five interconnected layers: the Infrastructure Layer, Data and Signal Layer, Cognitive Core Layer, Longitudinal Cognitive Operations Layer, and Strategic Stewardship Layer. Alongside these layers, the Tacit Signal Infrastructure captures implicit human cognitive signals and transforms them into machine-interpretable representations that can support adaptive reasoning and long-term calibration.

\subsection{L1: Infrastructure Layer}

The Infrastructure Layer provides the computational foundation for Longitudinal Cognitive Operations. It includes storage systems, networking, security mechanisms, scalable AI deployment environments, and the technical substrates required to support persistent human--AI cognitive systems.

This layer is important because tacit signal modelling cannot be supported by isolated tools or short-lived interaction sessions. It requires durable infrastructure capable of maintaining memory, securing sensitive cognitive data, enabling scalable computation, and supporting continuous system operation across time. Without this foundational layer, higher-level cognitive operations would remain fragmented and episodic.

The Infrastructure Layer also establishes the conditions for responsible system deployment. Because tacit signal infrastructure may involve sensitive data about professional behaviour, reasoning, interaction patterns, and expert judgement, issues of security, privacy, access control, and data stewardship must be treated as foundational rather than peripheral design concerns.

\subsection{L2: Data and Signal Layer}

The Data and Signal Layer integrates heterogeneous sources of information that may contribute to cognitive modelling. These sources may include structured data, interaction logs, contextual information, professional documents, task histories, decision traces, multimodal records, and synthetic cognitive data generated through simulation or AI-supported modelling.

This layer is where explicit data and potential tacit signals begin to enter the system. However, the framework does not assume that data directly reveals cognition. Interaction logs, behavioural traces, or contextual records are partial indicators that require interpretation. A hesitation, revision, change in tone, or repeated reframing may become meaningful only when situated within a broader semantic, temporal, and professional context.

The Tacit Signal Infrastructure is closely connected to this layer. It identifies weakly articulated signals within human activity, such as behavioural cues, expert intuition, contextual atmosphere, communicative shifts, interaction patterns, and signs of emerging tension. These signals are not treated as objective facts, but as candidates for interpretation that require further modelling, validation, and expert review.

\subsection{L3: Cognitive Core Layer}

The Cognitive Core Layer performs the central modelling and reasoning functions of the architecture. It supports perception, abstraction, reasoning, learning, and world modelling, enabling AI systems to move from raw data and signals toward structured cognitive interpretation.

Within this layer, explicit knowledge and tacit signals are organised into higher-level cognitive representations. For example, repeated conceptual inconsistency may be modelled as interpretive instability; changes in communicative rhythm may be interpreted as possible coordination breakdown; recurring hesitation around a decision pathway may suggest unresolved tension or uncertainty. These interpretations remain provisional and must be open to expert correction.

The Cognitive Core Layer therefore provides the bridge between signal detection and expert-level reasoning. It enables the system to construct models of coherence, tension, value conflict, semantic instability, expertise trajectory, and anticipatory risk. In this sense, the layer is not merely analytical; it is interpretive. It supports the transformation of weak signals into machine-interpretable structures that can be used for reflection, adaptation, and long-term calibration.

\subsection{L4: Longitudinal Cognitive Operations Layer}

The Longitudinal Cognitive Operations Layer supports the ongoing management of cognition across time. It includes long-term planning, memory-grounded reasoning, progress evaluation, cross-task knowledge consolidation, and the continuous calibration of cognitive models.

This layer is central to the paper's argument because tacit sensing is inherently longitudinal. Expert signals often become meaningful only through patterns of recurrence, change, intensification, resolution, or divergence over time. A single weak signal may be ambiguous, but a trajectory of signals can indicate emerging instability, coherence degradation, or expertise development.

Longitudinal Cognitive Operations therefore involve more than storing past interactions. They require the active maintenance of cognitive continuity: deciding what should be remembered, how memories should be retrieved, how semantic structures should evolve, how tacit signal interpretations should be revised, and how expert feedback should recalibrate the system. This layer transforms cognition from a sequence of isolated interactions into an evolving operational process.

\subsection{L5: Strategic Stewardship Layer}

The Strategic Stewardship Layer governs the long-term alignment, accountability, and societal orientation of cognitive infrastructures. It addresses alignment with human values, ethical principles, institutional responsibilities, and intergenerational objectives.

This layer is necessary because AI systems that model expert tacit sensing may influence professional judgement, organisational decisions, training pathways, and future planning. Such systems should not be governed only by technical performance metrics. They require stewardship mechanisms that consider who controls the infrastructure, whose expertise is represented, how interpretations are validated, how users can contest system outputs, and how long-term consequences are assessed.

Strategic stewardship also addresses the risk that tacit signal modelling may become a form of cognitive surveillance or managerial control. Because tacit signals may involve sensitive interpretations of behaviour, emotion, attention, uncertainty, trust, or collaboration, governance must ensure transparency, consent, accountability, and appropriate limits on use.

\subsection{Tacit Signal Infrastructure as a Cross-cutting Layer}

Although the architecture is organised into five layers, Tacit Signal Infrastructure functions as a cross-cutting mechanism that connects data capture, cognitive modelling, longitudinal operations, and governance. Its role is to transform implicit human cognitive signals into machine-interpretable representations while preserving the ambiguity, context-dependence, and contestability of expert judgement.

Tacit Signal Infrastructure may draw on behavioural cues, expert annotations, contextual atmosphere, interaction traces, reflective accounts, multimodal records, and longitudinal patterns. These inputs are not sufficient on their own. They must be interpreted through semantic structures, calibrated through expert feedback, and governed through ethical safeguards.

The purpose of Tacit Signal Infrastructure is therefore not to fully automate expert intuition. Rather, it supports the modelling of expert sensing as an evolving relationship between human judgement and AI-supported interpretation. It allows future AI systems to learn not only from explicit knowledge, but also from the weak signals through which experts anticipate change, recognise emerging tension, and maintain coherence in complex environments.

\subsection{Longitudinal Timeline and Future-oriented Adaptation}

The timeline at the bottom of Figure~\ref{fig:longitudinal-cog-operations} emphasises the temporal nature of the framework. Longitudinal Cognitive Operations extend from accumulated past experience, through present adaptive reasoning, toward future-oriented planning and legacy impact assessment.

This temporal orientation is important because the proposed framework does not treat cognition as a static state. Instead, cognition is understood as a trajectory. Past interactions, expert annotations, tacit signals, and semantic structures accumulate into memory; present reasoning draws on this memory to interpret current situations; and future planning uses these interpretations to guide adaptation, intervention, and long-horizon decision-making.

Overall, the architecture proposes that future AI systems should evolve from short-term response engines into persistent cognitive infrastructures. Such systems would be capable of continuous memory accumulation, tacit signal modelling, adaptive reasoning, and long-horizon human--AI coordination.

\section{Cognitive Operations Managers and Professional Capability}

The proposed framework implies a new class of professional capability. If next-generation AI systems are to model expert tacit sensing, then organisations will require people who can coordinate the relationship between explicit knowledge, tacit signal modelling, AI reasoning, expert validation, and ethical governance. This paper introduces the \textit{Cognitive Operations Manager} as a prototype role for describing this capability.

\begin{figure}[t]
    \centering
    \includegraphics[width=\linewidth]{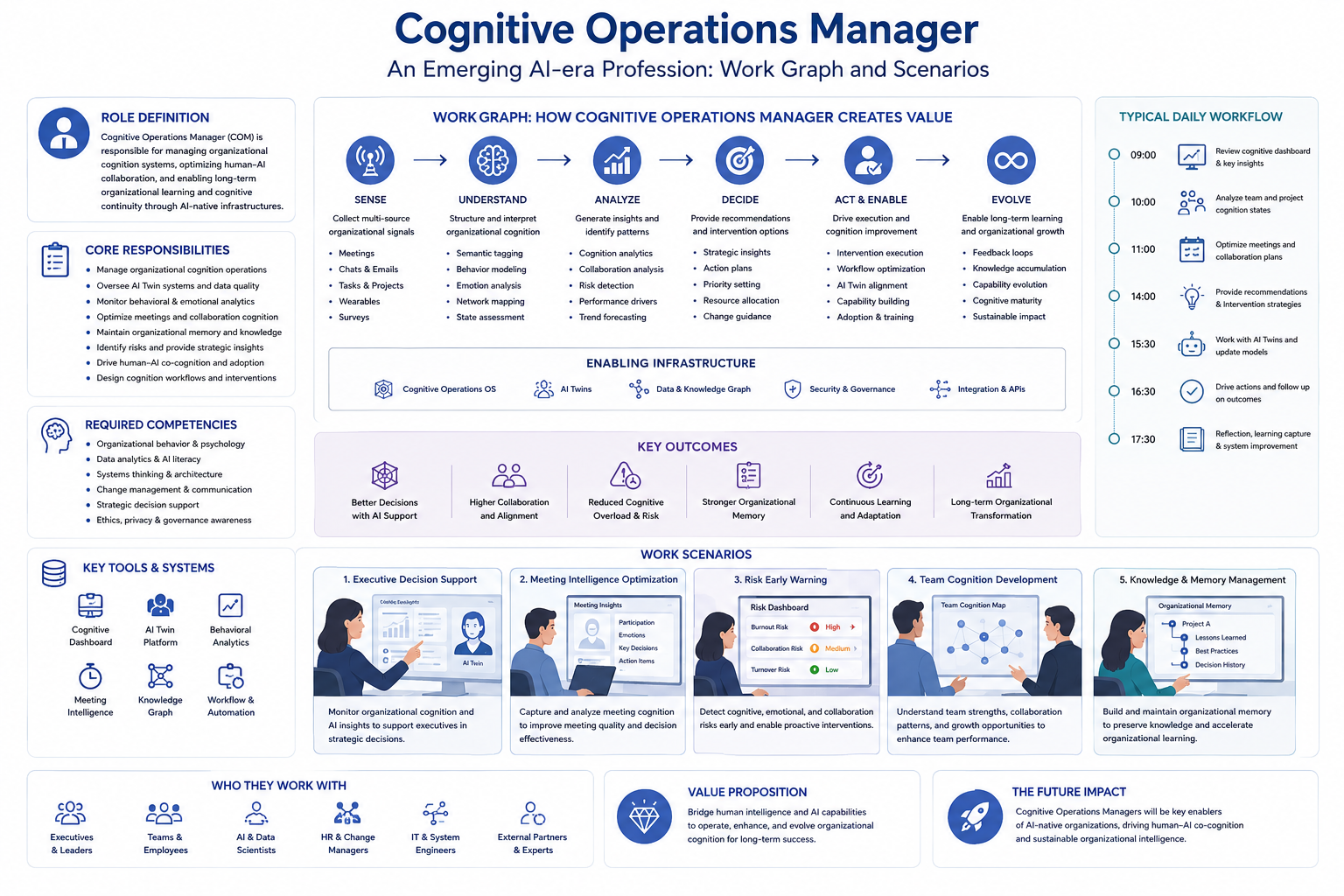}
    \caption{Cognitive Operations Manager: an emerging AI-era profession, work graph, and scenarios.}
    \label{fig:cognitive-operations-manager}
\end{figure}

Figure~\ref{fig:cognitive-operations-manager} illustrates the Cognitive Operations Manager as an emerging AI-era profession. The figure summarises the role's required competencies, core responsibilities, typical workflow, application scenarios, and value creation mechanisms. It positions the COM not as a conventional manager of people or tasks, but as a professional responsible for operating and governing cognitive infrastructures in which human expertise, AI systems, memory structures, semantic models, and tacit signals interact over time.

\subsection{Defining the Cognitive Operations Manager}

\begin{quote}
\textbf{Definition.} A Cognitive Operations Manager is an AI-native professional responsible for orchestrating persistent cognitive infrastructures, including memory systems, semantic structures, tacit signal layers, AI reasoning processes, expert feedback mechanisms, and governance safeguards.
\end{quote}

The COM is not proposed as a fixed job title or a fully established profession. Rather, it is a conceptual anchor for understanding an emerging capability space. As AI systems become more deeply involved in modelling expertise, organisations will need professionals who can ensure that these systems are meaningful, valid, ethical, and aligned with human judgement.

The COM differs from a prompt engineer, data analyst, product manager, or knowledge manager. Prompt engineers optimise interactions with AI systems. Data analysts interpret structured data. Product managers coordinate system development. Knowledge managers oversee organisational knowledge resources. The COM operates across these domains, but focuses specifically on the long-term operation of cognitive infrastructure.

\subsection{Responsibilities of the Cognitive Operations Manager}

The responsibilities of a COM span several interrelated forms of cognitive infrastructure work. A central task is coordinating \textit{tacit signal modelling}: working with domain experts to identify weak signals that matter in a given professional context and designing mechanisms through which those signals can be surfaced, interpreted, and reviewed without being oversimplified.

Because tacit signals only become meaningful within context, the COM also supports \textit{semantic modelling}. This includes maintaining domain ontologies, concept maps, value structures, and semantic relationships that connect weak signals to professional goals, histories, practices, and constraints. In this way, tacit signal modelling remains grounded in the domain knowledge that gives such signals meaning. The COM also oversees \textit{AI system calibration}. This involves reviewing whether AI-generated interpretations align with expert judgement, identifying reasoning drift, and preventing systems from overstating uncertain inferences. Such calibration is especially important because weak-signal interpretations should be treated as provisional hypotheses rather than definitive conclusions.

Another responsibility is maintaining \textit{expert feedback} loops. Since tacit signal interpretations must remain contestable, experts need opportunities to confirm, reject, refine, or reinterpret AI-generated patterns. These feedback mechanisms allow the cognitive infrastructure to evolve through ongoing human judgement rather than relying solely on automated inference. Finally, the COM contributes to \textit{ethical governance}. Tacit signal infrastructure may involve sensitive interpretations of professional behaviour, cognition, emotion, or team dynamics. The COM therefore helps establish appropriate safeguards around privacy, consent, transparency, accountability, contestability, and limits on use.

\subsection{Workflow and Value Creation}

The work of a Cognitive Operations Manager can be understood as a recurring operational cycle. A COM first identifies the cognitive domain or professional context in which tacit signals matter. They then work with domain experts to define relevant semantic structures, weak-signal categories, and validation criteria. During system operation, the COM monitors AI-generated interpretations, reviews emerging tacit signal patterns, coordinates expert feedback, and adjusts the cognitive infrastructure accordingly.

This workflow creates value in several ways. First, it improves the quality of AI-supported reasoning by ensuring that AI systems are calibrated against expert judgement rather than operating only on explicit data. Second, it supports earlier recognition of emerging risks, tensions, or incoherence by making weak signals more visible and reviewable. Third, it helps preserve human agency by ensuring that tacit signal interpretations remain contestable and governed. Fourth, it contributes to expertise development by turning tacit sensing into a reflective, teachable, and system-supported practice.

In this way, the COM functions as a mediator between human expertise and AI cognitive infrastructure. The role is valuable not because it replaces expert judgement, but because it helps organise the conditions under which expert judgement can be modelled, supported, and responsibly extended over time.

\subsection{Why This Role Becomes Necessary}

The need for this role emerges from the complexity of modelling expert cognition beyond explicit knowledge. Tacit signals are ambiguous, domain-specific, and temporally situated. They cannot be managed through technical automation alone. Nor can they be left entirely to individual experts without infrastructure support, because longitudinal patterns may exceed what any single person can track.

The COM therefore mediates between human expertise and AI infrastructure. This role ensures that AI systems remain grounded in expert judgement while also enabling expert cognition to be modelled, structured, and analysed over time.

This professional capability is especially important in domains where early recognition of weak signals matters, such as education, healthcare, design, engineering, organisational leadership, crisis management, creative production, and complex project environments.

\section{Cognitive Operations Research and Training Framework}

If Long-term Cognitive Operations represents a new operational paradigm, then it requires systematic research and training. Current AI education often focuses on tool use, prompt formulation, responsible AI awareness, and productivity enhancement. These competencies remain important, but they are insufficient for professionals who must design, evaluate, and operate AI systems that model expert tacit sensing over time.

This section introduces the \textit{Cognitive Operations Research and Training Framework} (CORTF) as a preliminary architecture for developing this capability.

\begin{quote}
\textbf{Definition.} The Cognitive Operations Research and Training Framework is a research and educational architecture for developing the capabilities required to design, evaluate, govern, and operate cognitive infrastructures that model expert cognition beyond explicit knowledge.
\end{quote}

CORTF connects research, infrastructure, training, and evaluation. Its purpose is not merely to teach people how to use AI systems, but to cultivate the ability to work with cognitive infrastructures responsibly and effectively.

\subsection{Research Layer}

The research layer investigates how expert tacit sensing can be identified, modelled, validated, and governed. Key research questions include:

\begin{itemize}
    \item How do experts perceive weak signals in different professional domains?
    \item Which tacit signals can be meaningfully captured through interaction traces, documents, multimodal data, or reflective accounts?
    \item How can AI systems distinguish meaningful weak signals from noise?
    \item How should expert feedback be incorporated into tacit signal modelling?
    \item How can longitudinal cognition models be evaluated ethically and empirically?
\end{itemize}

This research layer establishes the empirical and theoretical foundation for Tacit Signal Infrastructure.

\subsection{Infrastructure Layer}

The infrastructure layer provides environments for developing and studying cognitive operations. These may include semantic workspaces, AI memory systems, expert annotation tools, dashboard interfaces, AI Twin systems, multimodal data capture environments, and longitudinal cognition repositories.

The infrastructure layer is essential because tacit sensing cannot be studied only through abstract theory. It requires systems that allow researchers and practitioners to observe, model, validate, and revise interpretations over time.

\subsection{Training Layer}

The training layer develops the capabilities required for Cognitive Operations Managers and related AI-native professionals. Core training areas include:

\begin{itemize}
    \item \textit{Tacit signal recognition}: learning to identify weak signals, emerging tensions, coherence shifts, and anticipatory instability.
    \item \textit{Semantic orchestration}: organising concepts, categories, and relationships so that tacit signals can be interpreted in context.
    \item \textit{Memory curation}: managing what AI systems remember, retrieve, revise, or forget.
    \item \textit{Reasoning calibration}: evaluating AI-supported interpretations and identifying reasoning drift.
    \item \textit{Expert feedback design}: creating mechanisms through which human experts can validate and contest AI interpretations.
    \item \textit{Cognitive ethics}: addressing privacy, consent, surveillance, dependency, accountability, and autonomy.
\end{itemize}

This training agenda shifts AI education beyond prompt literacy toward cognitive infrastructure capability.

\subsection{Evaluation Layer}

The evaluation layer assesses whether cognitive operations are effective, ethical, and developmentally valuable. Conventional evaluation metrics such as task completion, speed, or output accuracy are insufficient. Long-term Cognitive Operations require evaluation at multiple levels.

Possible evaluation dimensions include:

\begin{itemize}
    \item \textit{Memory quality}: whether retained information remains accurate, relevant, and appropriately contextualised.
    \item \textit{Semantic coherence}: whether concepts and relationships remain consistent and meaningful over time.
    \item \textit{Tacit signal validity}: whether detected weak signals correspond to expert-recognised patterns.
    \item \textit{Reasoning accountability}: whether AI interpretations can be explained, reviewed, and corrected.
    \item \textit{Expertise development}: whether the system supports human learning and judgement rather than replacing it.
    \item \textit{Governance quality}: whether privacy, consent, contestability, and ethical safeguards are maintained.
\end{itemize}

Through these layers, CORTF provides a research and training foundation for the next generation of AI cognitive systems.

\section{Discussion}

The framework proposed in this paper is intentionally forward-looking. It does not claim that AI systems can already model expert tacit sensing reliably, nor that all tacit cognition should be formalised. Rather, it argues that as AI systems become increasingly capable of handling explicit knowledge, the frontier of expert cognition lies in tacit sensing, longitudinal judgement, and weak-signal interpretation. Modelling this layer requires new infrastructures, operations, professions, and safeguards.

\subsection{Avoiding the Over-formalisation of Tacit Knowledge}

A central risk is the over-formalisation of tacit knowledge. Tacit sensing is valuable partly because it is situated, flexible, and context-sensitive. If AI systems force tacit signals into rigid categories, they may distort the very expertise they aim to model.

Tacit Signal Infrastructure should therefore avoid treating tacit cognition as fully capturable data. Instead, it should support partial representation, expert reflection, and iterative validation. The goal is not to convert tacit judgement into fixed rules, but to create infrastructures through which tacit signals can be surfaced, discussed, and refined.

\subsection{False Weak-Signal Detection}

Another risk concerns false weak-signal detection. AI systems may identify patterns that appear meaningful but are actually noise, bias, or artefacts of data collection. Because weak signals are ambiguous by nature, there is a danger that AI-generated interpretations could be over-trusted.

For this reason, tacit signal modelling must include uncertainty representation, expert review, and contestability. AI systems should present weak-signal interpretations as hypotheses rather than conclusions.

\subsection{Cognitive Surveillance and Power}

Tacit signal infrastructure may involve sensitive interpretations of professional behaviour, cognition, emotion, or collaboration. In organisational contexts, such systems could become tools of surveillance if used to monitor workers' cognitive states, emotional engagement, or perceived instability.

This risk is especially serious because tacit signals are not neutral facts. They are interpretations. Governance structures must therefore define who can access tacit signal models, how interpretations may be used, how individuals can contest them, and what forms of monitoring are inappropriate.

\subsection{Human Expertise and Cognitive Dependency}

AI systems that model expert sensing could support professional development, but they could also create cognitive dependency. If professionals rely too heavily on AI-generated interpretations, they may lose opportunities to develop their own tacit judgement.

The design goal should therefore be augmentation rather than substitution. Tacit Signal Infrastructure should help experts notice, reflect, and learn. It should not replace human responsibility for judgement.

\subsection{Domain Specificity}

Tacit signals differ across domains. The weak signals that matter in engineering design may differ from those in education, clinical care, organisational leadership, or creative practice. Therefore, Tacit Signal Infrastructure cannot rely on a universal taxonomy of expert sensing.Future systems must be domain-sensitive and co-developed with experts. This also means that Cognitive Operations Managers will need domain knowledge, not only technical AI skills.

\section{Future Research Directions}

The proposed framework opens several directions for future research. Empirical work is needed to understand how experts perceive, describe, and validate tacit signals across domains. Methods such as interviews, think-aloud protocols, reflective diaries, workplace observations, and expert annotation of professional episodes could help identify the weak signals that practitioners treat as meaningful in context.

A second direction concerns representation. Future research should examine how tacit signals can be modelled without over-formalising them or reducing them to rigid categories. This may involve flexible coding schemes, uncertainty-aware models, and expert-in-the-loop validation processes that preserve the situated and provisional nature of tacit judgement. Technical research is also needed to develop longitudinal architectures for tacit signal modelling. Such architectures would need to connect interaction traces, semantic ontologies, memory systems, weak-signal detection, and expert feedback across time. This is essential if AI systems are to model not only isolated signals, but also their recurrence, trajectory, and changing significance.

Evaluation remains another major challenge. Future work should develop frameworks for assessing whether AI systems can meaningfully support expert sensing. Relevant dimensions may include signal validity, expert trust, reasoning accountability, uncertainty communication, and the impact of such systems on professional judgement over time. Governance research is equally important. Tacit signal systems may involve sensitive interpretations of behaviour, cognition, emotion, collaboration, or professional capability. Future work should therefore address privacy, consent, contestability, accountability, and responsible limits on use, particularly in workplaces, education, healthcare, and other high-stakes domains.

Finally, education and workforce research should investigate how Cognitive Operations Managers and related professional roles might be trained. Such training would need to combine AI literacy, domain expertise, tacit signal interpretation, semantic modelling, reasoning governance, and ethical judgement. This would help shift AI education from tool use alone toward the development of cognitive infrastructure capability.

\section{Conclusion}

This paper has argued that the next generation of AI systems should move beyond explicit knowledge processing toward the modelling of expert tacit sensing over time. Current generative AI systems are increasingly effective at retrieving, summarising, organising, and operationalising explicit knowledge. However, high-level expertise also depends on tacit sensing: the ability to perceive weak signals, recognise emerging tensions, detect coherence degradation, and anticipate instability before formal indicators appear.

To address this gap, the paper introduced \textit{Tacit Signal Infrastructure} as a cognitive infrastructure layer for capturing, structuring, interpreting, and validating expert tacit signals over time. It further defined \textit{Long-term Cognitive Operations} as the operational paradigm required to maintain and govern AI systems that model expert cognition beyond explicit knowledge.

The paper also proposed the \textit{Cognitive Operations Manager} as a prototype professional role responsible for coordinating tacit signal capture, semantic modelling, AI system calibration, expert feedback, and ethical governance. Finally, it introduced the \textit{Cognitive Operations Research and Training Framework} as a research and educational architecture for developing the capabilities required to design, evaluate, and operate next-generation AI cognitive systems.

The central contribution of this paper is to reposition cognition as an infrastructural and operational domain in the AI era. If future AI systems are to support expert cognition meaningfully, they must not only process what experts know explicitly. They must also develop ways to model how experts sense, interpret, and anticipate change across time. This requires not only better models, but new cognitive infrastructures, operational practices, professional capabilities, and ethical safeguards.

\subsection*{Copyright Notice}

\noindent
\textcopyright\ 2026 Annie Yihong Yuan. All rights reserved. AAll figures, diagrams, interface examples, and visual materials in this paper are original works of the author unless otherwise stated.


\bibliographystyle{unsrtnat}
\bibliography{references}  






\end{document}